\documentclass[12pt, a4paper]{article}
\usepackage[top=2cm, bottom=2cm, left=1.5cm, right=1.5cm, includefoot]{geometry}
\usepackage{multicol}
\usepackage{graphicx}
\usepackage{amsmath}
\usepackage{amssymb}
\usepackage{amsfonts}
\usepackage{color}
\usepackage{wrapfig}
\usepackage{eucal}
\usepackage{hhline}
\usepackage{threeparttable}
\usepackage{supertabular}
\usepackage{multirow}
\usepackage{tabularx}
\usepackage{siunitx}
\usepackage{float}
\usepackage[export]{adjustbox}
\usepackage{soul}
\usepackage[version=4]{mhchem}
\usepackage{array}
\usepackage[colorlinks=true,
    citecolor=green!70!black,   
    linkcolor=blue,             
    urlcolor=red,               
    filecolor=magenta           
]{hyperref}
\usepackage{comment}
\usepackage{bm}
\usepackage{booktabs}
\usepackage{makecell}
\usepackage{tikz}
\usetikzlibrary{arrows.meta, positioning, shapes.geometric}
\usepackage{amsthm}

\usepackage{makeidx}
\makeindex 

\usepackage[numbers, sort&compress]{natbib}
\usepackage{times}      
\usepackage{helvet}     
\usepackage{courier}    
\usepackage{mathptmx}   

\usepackage{titlesec}
\titleformat{\section}
    {\Large\bfseries\sffamily}
    {\thesection}{1em}{}
\titleformat{\subsection}
    {\large\bfseries}
    {\thesubsection}{1em}{}
\titleformat{\paragraph}
    {\normalsize\bfseries}
    {\theparagraph}{1em}{}
\titlespacing*{\paragraph}{0pt}{3.25ex plus 1ex minus .2ex}{1em}

\newcommand{\keywords}[1]{%
  \par\vspace{10pt}\noindent
  \hangindent=4em\hangafter=1
  \textbf{Keywords: } #1\par
}
\begin{document}

\title{\bfseries\sffamily Isomorphic Emergence of Lorentz and Gauge Symmetries\\[0.3em]
\textemdash A Constructive Interpretation Based on Continuum Mechanics}
\author{
    Ke-Xia Jiang\thanks{Email: kexiajiang@126.com} \\
    Department of Physics, Engineering University of PAP, Xi'an 710086, P. R. China}
\maketitle

\begin{abstract}
Lorentz symmetry and gauge symmetry constitute the mathematical cornerstones of modern physics, yet their ultimate physical origins remain elusive. From the standpoint of a constructive interpretation, this paper demonstrates that both symmetry structures can emerge isomorphically from a unified classical source: dynamical constraints on wave-packet excitations in a continuous elastic substrate medium (SM). Neither symmetry is posited as fundamental, nor does their emergence rely on quantization. Three core results are established. First, taking the transverse wave speed of a homogeneous, isotropic SM as the benchmark under a conventionalist synchronization scheme, Minkowski-type spacetime arises isomorphically, with Lorentz transformations as effective coordinate transformations between inertial frames. This reinterprets the Michelson-Morley null result: observers are composite wave-packet excitations, and medium-induced kinematic corrections are encoded in their measurement frameworks. Second, for wave packets endowed with $SU(N)$ intrinsic symmetry, the requirement of identity invariance drives the SM to spontaneously generate gauge fields. This requires invariant equivalence classes of intrinsic states under propagation, yielding gauge structures isomorphic to Yang-Mills theory. The gauge group $SU(N)$ is uniquely determined by the number of stable normal modes supported by the medium. Third, through the same conventionalist measurement protocols, inhomogeneous distributions of the SM emerge isomorphically as curved spacetime geometry, governed by Einstein-type field equations from variational extremization of the medium's deformation free energy. This paper advances a unified interpretive account of the physical origin underlying the mathematical structures of both Relativity and the Standard Model.
\end{abstract}




\section{Introduction}
\label{sec:introduction}

Although theories of the aether had dominated physical conceptions of space and electromagnetic wave propagation for nearly two centuries~\cite{Whittaker1910}, the null result of the Michelson-Morley experiment on the relative motion of the Earth and the luminiferous aether~\cite{MichelsonMorley1887} foreshadowed a profound crisis for the aether hypothesis. In the early twentieth century, the formulation of special relativity (SR)~\cite{Einstein1905b} marked the demise of the luminiferous aether as an absolute reference frame and carrier of electromagnetic waves, rendering the aether a superfluous concept in physics. At the time SR was formulated, electromagnetic waves stood as the sole known wave phenomenon that propagated without a material medium. Since then, Minkowski spacetime $\mathbb{M}^{1,3}$---an \textit{a priori}, immaterial geometric arena---has served as the foundational spacetime structure of modern physics. Against this background, the gauge principle has achieved remarkable success: from Weyl's original gauge conception~\cite{Weyl1918_Gravitation} to Yang-Mills theory~\cite{YangMills1954}, and further to the establishment of electroweak unification~\cite{Weinberg1967} and the discovery of asymptotic freedom in quantum chromodynamics~\cite{GrossWilczek1973,Politzer1973}, the mathematical structure of gauge invariance has proven capable of precisely describing the dynamics of the three fundamental interactions---electromagnetic, weak, and strong.

However, gravity has yet to be incorporated into the gauge framework of the Standard Model; general relativity (GR) interprets gravity as the geometry of spacetime, a perspective that is fundamentally distinct at the methodological level from the ontological premises of quantum field theory. Beneath this lies a deeper question: is gauge symmetry a fundamental law of nature, or merely an effective manifestation of deeper physics in the low-energy limit? If the latter proves true, the choice of the gauge group $SU(3) \times SU(2) \times U(1)$ would cease to be arbitrary, but would instead be determined by underlying dynamical constraints. Likewise, the local form of Lorentz symmetry should not be regarded as an \textit{a priori} geometric postulate; it ought to have a traceable physical origin.

When we turn to condensed matter physics, an intriguing parallel emerges: gauge structures, long considered a hallmark of high-energy physics, recurrently appear in the effective descriptions of these systems. The fractional quantum Hall effect~\cite{Tsui1982,Laughlin1983} demonstrates that physical laws can ``emerge'' from a simpler microscopic substrate; these emergent phenomena often exhibit symmetries and topological properties absent from their microscopic constituents, thereby broadening our understanding of gauge theory. The phase structure of the Aharonov-Bohm effect can also be interpreted within a classical framework~\cite{Kholmetskii2013}, while synthetic gauge fields can be engineered and manipulated in cold-atom systems~\cite{Goldman2014} and photonic systems~\cite{Song2025,Leykam2026}. These condensed-matter realizations suggest that gauge structures need not be intrinsic attributes of spacetime itself, but may instead constitute universal features of collective excitation dynamics.

 Moreover, in classical generalized continuum mechanics (GCM), gauge structures are equally deeply embedded at the heart of the theoretical framework. The pioneering work of the Cosserat brothers on deformable media~\cite{Cosserat1909} established the foundations for describing the independent rotational degrees of freedom of material microstructures. In the 1950s, the equivalence between crystal dislocation density and the torsion tensor in Riemann-Cartan geometry~\cite{Kondo1952,Bilby1955} was established, initiating a research tradition that describes defects in the language of differential geometry. Subsequently, a gauge theory of dislocations and disclinations was systematically developed~\cite{Kadic1983} and extended to amorphous solids~\cite{DereliVercin1987}; this framework was further systematized, enabling the application of Yang-Mills theory to the dynamical analysis of defective solids~\cite{EdelenLagoudas1988}, and was extended to metric-torsion gauge theories of continuum line defects~\cite{Vercin1990} and to gauge theories of Cosserat surfaces~\cite{Badur1993}. In the realm of fluid dynamics and dynamic plasticity, the rotational degrees of freedom of microscopic vortices in turbulence can likewise be characterized by the torsion tensor~\cite{Peshkov2019}, while the spacetime gauge theory of dynamic plasticity~\cite{Kumar2024} further extends the scope of gauge structures in continuous media. In these classical systems, gauge structures do not reside in an \textit{a priori} geometric spacetime, but rather emerge from the collective dynamics of concrete physical media, such as crystals, elastic continua, and fluids. Similarly, there exists a precedent for reinterpreting electrodynamics through continuum models: the dynamical equations of linear elastic media can be formally mapped onto Maxwell's equations~\cite{Christov2006,Christov2011}, and systematic investigations have examined the foundations of electrodynamics from the viewpoint of rational continuum mechanics (CM)~\cite{Mueller2023}. These considerations indicate that the mathematical framework of gauge theory may be rooted in classical dynamics and collective behavior, existing independently of quantization. As Witten~\cite{Witten2018_NaturePhys} has emphasized, gauge symmetry may well emerge from more fundamental physics.

The geometric language employed in GCM---frame fields, connections, and curvature tensors~\cite{deWit1981,Eringen1999}---shares core conceptual elements with the geometric description of gravity in GR. The linear correspondence between crystal defects and Riemann--Christoffel curvature as well as Cartan torsion~\cite{deWit1981}, the interpretation of static point-particle solutions in three-dimensional gravity as linear defects in solids~\cite{KatanaevVolovich1992}, and the derivation of Einstein-type field equations for defective continua from a variational principle applied to curvature-quadratic terms in the free-energy functional~\cite{Katanaev2005} all suggest that the similarity between the geometric languages of GR and CM may have a deeper origin.

Motivated by the above considerations, this paper adopts, within the framework of classical continuum theory,  a \textit{constructive interpretation}: without presupposing spacetime or gauge symmetry as fundamental, we start from a set of simpler dynamical assumptions and mathematically reproduce these structures, seeking a unified physical origin for the geometric structures of gauge fields and gravity. The specific development proceeds as follows. First, a homogeneous and isotropic substrate medium (SM), together with wave packets excited thereon, gives rise---under a conventionalist time-synchronization scheme---to emergent Minkowski-type spacetime and Lorentz symmetry. Second, the dynamical requirement of identity invariance for the internal degrees of freedom of wave packets, through their interaction with the SM, induces the necessary response of the medium---namely, the generation of gauge fields---whose mathematical structure is isomorphic to that of Yang-Mills theory. Third, the inhomogeneous distribution of the SM, through the same conventionalist operation, emerges as curved spacetime geometry, whose dynamics are governed by Einstein-type field equations. The unifying power of this framework lies not in subsuming all interactions into a single group-theoretic structure, but in tracing them back to the same physical source: the SM and its wave-packet excitations.

The remainder of this paper is organized as follows. Sec.\ref{sec:ether} reviews the legacy and lessons of aether theory and clarifies the methodological basis of the present work; Sec.\ref{sec:wavepacket} establishes the internal symmetry of multi-component wave packets; Sec.\ref{sec:continuum} outlines the gauge-geometric structure of GCM; Secs.~\ref{sec:spacetime}, \ref{sec:gauge}, and \ref{sec:gravity_emergence} discuss the emergent mechanisms of spacetime, gauge fields, and gravity, respectively; and Sec.\ref{sec:prospects} presents discussion and conclusions.

\section{The Legacy and Lessons of Aether Theory}
\label{sec:ether}

We begin by reviewing the historical concept of the ``aether,'' which serves to clarify both the conceptual kinship and the distinction between the framework proposed here and historical aether theories, to reveal the problems inherent in those theories, and thereby to explain why the present framework can circumvent these difficulties.

The term ``aether'' originates from ancient Greek philosophy, where it denoted the fifth element permeating cosmic space. From the seventeenth to the nineteenth century, with the development of wave optics, the aether was conceived as a medium for the propagation of light; at that time all known wave phenomena required a material medium, making this a natural and reasonable assumption. For a detailed historical account, one may consult Whittaker's treatise~\cite{Whittaker1910}, which systematically traces the development of aether and electricity theories from the age of Descartes to the close of the nineteenth century, documenting the pioneering efforts of that era.

\subsection{The Dilemma of Aether Theory}
\label{sec:ether_dilemma}

The core assumptions of aether theory rested on two tenets. First, space was filled by a stationary medium that furnished an absolute rest frame and served as the carrier of electromagnetic waves. Second, material bodies moved through this medium, analogous to objects moving through a fluid. This picture implied that bodies and the medium were two independently identifiable kinds of entities---what we shall call the \textit{dualistic hypothesis}.

As Einstein~\cite{Einstein1952} later emphasized, in accordance with classical mechanics and SR, ``space (space-time) has an existence independent of matter or field''; if matter and field were removed, ``inertial-space or, more accurately, this space together with the associated time remains behind.'' The four-dimensional structure was ``thought of as being the carrier of matter and of the field''---a conception that Einstein himself identified as a four-dimensional analogue of Lorentz's rigid three-dimensional aether. In his 1920 Leiden lecture, Einstein~\cite{Einstein1920} characterized this mechanical aether as possessing the properties of ponderable matter, consisting of parts that could be traced in time and to which the notion of motion could be ascribed.

It is worth noting that the idea of an absolute frame of reference in physics did not originate with the electromagnetic aether of the nineteenth century. The ``absolute space'' tacitly assumed in Newtonian mechanics played an analogous role: it entered the laws of motion as a physical reality, yet remained unaffected by the distribution of matter. From Descartes to Lorentz, the aether underwent successive transformations---from vortex particles to a quasi-rigid solid---yet its core dualistic structure persisted, namely, the coexistence of an independent, unobservable medium and identifiable bodies moving within it.

We maintain that it is precisely this dualistic hypothesis---treating the medium as an absolute background independent of bodies---that engendered the dilemma of aether theory. If bodies can be identified and tracked independently of the medium, then their motion relative to the medium is, in principle, detectable. The Michelson-Morley experiment~\cite{MichelsonMorley1887} was precisely a test of this detectability, and its null result signaled the failure of this dualistic picture.

\subsection{Philosophical Presuppositions of Spacetime}
\label{sec:spacetime_presupposition}

Be it Newton's absolute spacetime or the Minkowski spacetime of SR, both rest on a common premise: that space can exist as a background arena independently of matter, permitting the existence of completely empty regions. Although GR dynamically couples spacetime to matter and identifies the gravitational field with spacetime geometry, its field equations nevertheless admit vacuum solutions in the absence of matter.

The Galilean coordinate system of Newtonian absolute spacetime constitutes a purely mathematical idealization---an absolute frame that is not physically operable, for no definition of simultaneity independent of physical signals exists, and all such signals propagate at finite speed. This does not imply, however, that Newtonian absolute spacetime is devoid of meaning; it remains a limiting idealization available as a reference framework. The relativistic conception of spacetime overcame precisely this limitation by providing an operational definition of time through the synchronization of light signals~\cite{Einstein1905b}. Yet, by postulating the constancy of the speed of light, Relativity severed light from any material carrier, thereby circumventing the problem of a material medium at the operational level.

It should be emphasized that the task of physics with respect to spacetime is not to determine its ultimate essence, but to provide a mathematical description thereof. Countless experiments over the past century have confirmed the validity of the relativistic mathematical description of spacetime; the aim of the present paper is to understand and interpret this description through a new theoretical framework.

\subsection{Einstein's Ambivalence}
\label{sec:einstein_ambivalence}

In any examination of aether theory, Einstein's stance is particularly representative, as well as notably cautious and complex, clearly reflecting the depth of this dualistic dilemma.

In 1905, in founding SR~\cite{Einstein1905b}, Einstein, proceeding from the principle of the constancy of the speed of light and the principle of relativity, resolutely abandoned the hypothesis of the ``luminiferous aether,'' maintaining that the electromagnetic field constitutes a form of physical reality capable of propagating through vacuum as waves without any medium as a carrier. At this stage, Einstein adopted the view that space constitutes an empty background, devoid of intrinsic physical properties.

However, in his celebrated 1920 Leiden lecture entitled ``Ether and the Theory of Relativity''~\cite{Einstein1920}, Einstein's attitude underwent a marked shift. He explicitly stated that, according to GR, space is endowed with physical qualities, and that in this sense an aether exists. He distinguished two conceptions of the aether: first, the mechanical aether of the nineteenth century; second, the aether of GR, which is space itself considered as endowed with physical properties. His position had now turned toward the identification of space with the field. Yet he imposed a crucial restriction on this new aether: one must not attribute to it the properties characteristic of a ponderable medium, conceive of it as consisting of parts that can be tracked through time, nor ascribe to it the notion of motion. This indicates that he sought a delicate balance between ascribing physical properties to space and avoiding a return to the mechanical aether.

In 1938, in \textit{The Evolution of Physics}, co-authored with Infeld~\cite{EinsteinInfeld1938}, Einstein reformulated his position once again, emphasizing that the physical reality of the field can exist independently of the name ``aether.'' This seemingly contradictory formulation reflects his persistent vacillation on the question of whether the field is independent of space.

In 1952, in the appendix ``Relativity and the Problem of Space'' to the fifteenth edition of \textit{Relativity: The Special and the General Theory}~\cite{Einstein1952}, Einstein offered his final assessment: spacetime does not claim an existence independent of the actual objects of the physical world, but is merely a structural quality of the field; physical objects are not in space, but these objects have spatial extension. By then, he had explicitly abandoned the notion of ``empty space,'' yet failed to clarify the physical essence of this spatial extension.

Einstein's vacillation reveals a fundamental problem: he never fully resolved the dualistic dilemma of space versus field. On the one hand, he maintained in SR that the electromagnetic field requires no medium; on the other hand, he acknowledged in GR that the gravitational field constitutes a physical property of spacetime itself. This very inconsistency indicates that the ontological status of the electromagnetic and gravitational fields is not on an equal footing within Einstein's framework: the gravitational field is geometrized, whereas the electromagnetic field is still regarded as a field residing in space.

These analyses jointly indicate that, although Einstein pioneered the idea of identifying space with the field, he did not thoroughly eliminate the dualistic separation between field and space, or between gravity and electromagnetism. This contradiction remains an unresolved problem in physics. The mainstream view has inherited this basic dualistic stance: spacetime constitutes the geometrized expression of gravity, whereas the electromagnetic field is regarded as a field superimposed upon spacetime.

Earman and Norton~\cite{EarmanNorton1987} argue that if one insists on spacetime substantivalism, treating spacetime as a ``stage'' independent of the field, GR faces a threat of indeterminism. In the context of quantum gravity, this problem assumes a sharper form: if the ontology of a fundamental theory does not include spacetime, the theory faces the threat of ``empirical incoherence,'' because all observation depends on localized entities in spacetime; how such observable localized entities can emerge becomes a central question that any theory of quantum gravity must address~\cite{HuggettWuthrich2013}. Romero, following a similar line of thought, proposes ``event substantivalism,'' arguing that spacetime as an entity emerges from more fundamental timeless and spaceless ``pre-geometric'' entities~\cite{Romero2017}, offering another philosophical avenue for questioning the primitive substantiality of spacetime.

\subsection{Conventionalism}
\label{sec:conventionalism}

Having analyzed the dualistic dilemma, we now turn to conventionalism. Between 1898 and 1902, Poincar\'e~\cite{Poincare1898,Poincare1902} argued that geometrical axioms and the measurement of time are essentially matters of ``convention,'' adopted so as to render the formulation of the laws of nature as simple as possible. This implies that seemingly \textit{a priori} necessary spacetime structures actually contain an irreducible element of free choice.

In his \textit{Philosophie der Raum-Zeit-Lehre} (1928), Reichenbach~\cite{Reichenbach1958} rendered this idea more concrete. He maintained that the definition of simultaneity in SR circumvents the direct measurement of the one-way speed of light. The fundamental reason lies in an unavoidable circularity: to measure the one-way speed of light, one must first synchronize clocks at two spatially separated locations; yet to synchronize such clocks, one must presuppose a value for the one-way speed of light. Specifically, SR employs light signals to synchronize clocks at locations A and B according to the convention $t_B = t_A + \varepsilon(t'_A - t_A)$, where $\varepsilon = 1/2$. This particular choice is equivalent to the conventional postulate that the one-way speed of light is isotropic and constant.

However, one could in principle adopt a nonstandard synchronization convention with $\varepsilon \neq 1/2$, in which case the dynamical equations would assume a more complicated form. For example, in an inertial frame with $\varepsilon \neq 1/2$, additional terms analogous to Coriolis and centrifugal forces appear in the dynamical equations~\cite{Ohanian2004}. Although conventionalists regard these terms as mere coordinate artifacts rather than genuine physical effects~\cite{Macdonald2005}, the choice $\varepsilon = 1/2$ does indeed preserve the dynamical equations in their simplest form. This indicates that the choice $\varepsilon = 1/2$ is not a purely arbitrary philosophical decision, but a coherent outcome of the interplay between physical theory and the definition of inertial frames: it is at once a convention chosen for descriptive simplicity and a constraint imposed by dynamical consistency. The philosophical interpretation of this dual nature remains a matter of ongoing debate; see \cite{Grunbaum1973,Malament1985,WeatherallManchak2014,Duerr2022}.

The above discussion presupposes a synchronization scheme for light signals in vacuum. It is noteworthy that the same conventionalist analysis remains valid even if one does not exclude the existence of a material medium for light. In this case, the state of motion of the medium relative to an inertial frame is in principle undetectable; any potential anisotropy of the signal speed is absorbed into the very definition of local time and spatial coordinates. Different local inertial frames preserve the covariance of spacetime structure by adopting the same conventional value for the speed of light, $c$.

Thus, we take the central lesson of conventionalism to be: when the constancy of the speed of light is incorporated into the definition of spacetime by convention, the possible effects of the absolute motion of a medium are systematically screened from the observational level. This does not mean that the medium does not exist, but rather that the medium is encoded within the very operational procedures through which we construct our observational framework. This idea constitutes the methodological core of the present paper and will be elaborated through concrete physical models and gauge structures in Secs.~\ref{sec:spacetime} and \ref{sec:gauge}.

\subsection{The Framework of the Present Work}
\label{sec:our_approach}

The aim of this paper is to offer a \textit{constructive interpretation}: to show that the mathematical structures of the Standard Model and of GR can be isomorphically traced back to one and the same physical source, namely wave-packet excitations in the SM. We do not attempt to establish a new physical theory; rather, we aim to present an interpretative framework in which the gauge structures of the Standard Model and the geometric structures of GR are understood as effective descriptions emergent from the dynamics of wave packets in the SM.

The fundamental postulates adopted herein are as follows. First, physical space is fully filled by the SM, and no void space free of the SM exists. Second, all familiar forms of ``matter", including electrons, protons, photons and even observers themselves, are wave-packet excitations supported on the SM. Third, wave packets carry intrinsic internal degrees of freedom governed by symmetry rules.

A core corollary of these postulates is that observers can neither step out of the SM to detect the SM itself, nor mark any reference object as independent of the SM. The SM cannot, in principle, be detected by the wave-packet excitations residing upon it; this is not a flaw of the theory, but an inevitable logical consequence inherent to the framework. Just as water waves can reflect macroscopic properties of water yet cannot reveal the underlying ontology of water molecules, measurement apparatuses composed of wave packets likewise fail to directly probe the SM that generates these wave packets. From this viewpoint, the null result of the Michelson-Morley experiment admits an alternative interpretation: the Earth itself is a complex assembly of wave-packet excitations, and all kinematic effects arising from its motion are encoded within the observational framework.

Our ontological stance does not regard ``spacetime'' and the SM as interchangeable labels. Instead, we argue that the spacetime geometry we perceive is an effective emergent description of SM dynamics under specific observational operations, rather than an \textit{a priori} entity. Specifically, homogeneous distributions of the SM lead to the isomorphic emergence of flat spacetime geometry, while inhomogeneous SM distributions manifest as curved spacetime geometry whose dynamics are governed by Einstein-type field equations. In this sense, spacetime and the SM are not two separate entities that need to be reduced to one another; rather, the SM constitutes the fundamental physical reality, while spacetime geometry provides an effective emergent description of that reality under specific observational operations.

Accordingly, asking which one is more fundamental between ``spacetime'' and the SM amounts to a pseudo-question. Raising the question of what space is prior to being filled by the SM falls into the trap of dualism that separates spacetime and the SM. This paper does not seek to explore the microscopic structure underlying the SM; instead, taking the existence of the SM as a working premise, we systematically investigate the feasibility of the proposed theoretical framework.

\section{Wave Packets with Intrinsic Degrees of Freedom}
\label{sec:wavepacket}

The Galilean coordinate system of Newtonian absolute spacetime constitutes a purely mathematical idealization ---an absolute frame that can serve as a limiting reference. We first construct the wave-packet description within such coordinates and subsequently extend the formalism to the general case.

\subsection{Scalar Wave Packets}
\label{sec:scalar_wavepacket}

Consider transverse wave excitations of the SM possessing a fixed polarization direction. The motion of the corresponding wave packet may be described by a scalar field wavefunction $\psi$. Let the velocity field take the form $\bm{v}(\bm{x},t) = \psi(\bm{x},t) \hat{\bm{\epsilon}}_0$, where $\hat{\bm{\epsilon}}_0$ denotes the oscillation direction. The wavefunction obeys the wave equation
\begin{equation}
\label{eq:klein_gordon_scalar}
L_0 \psi \equiv \left[ \frac{\partial^2}{\partial t^2} - c^2 \nabla^2 + \omega_0^2 \right] \psi = 0.
\end{equation}
Here, $c = \sqrt{\mu/\rho}$ denotes the transverse wave speed, $\mu$ the shear modulus, $\rho$ the mass density of the medium, and $\omega_0$ the intrinsic eigenfrequency. We interpret $|\psi(\bm{x},t)|^2$ as the relative energy density, with the wavefunction $\psi(\bm{x},t)$ satisfying energy normalization.

The velocity field formed by superposing multiple oscillation modes constitutes a scalar wave packet, whose integral representation for continuous modes reads
\begin{equation}
\label{eq:complex_wavefunction}
\Psi(\bm{x},t) = \int d^3k \, f(\bm{k}) e^{i(\bm{k}\cdot\bm{x} - \omega t)},
\end{equation}
where $f(\bm{k})$ denotes the complex amplitude distribution. The dispersion relation for the oscillation modes is given by
\begin{equation}
\label{eq:dispersion_scalar}
\omega^2 = c^2 k^2 + \omega_0^2.
\end{equation}

The scalar wave packet described above characterizes wave excitations restricted to a single fixed polarization direction and serves as a special case of the multi-component wave packets introduced below. When generalizing to $N$ normal modes, these modes are treated as abstract, linearly independent oscillations residing in the system's state space, rather than as oscillations confined to a single fixed polarization in physical space. In a dissipation-free elastic SM, energy conservation guarantees that localized excitations do not decay through dissipation; all modes with distinct wavevectors share an identical dispersion relation within a homogeneous and isotropic SM.

\subsection{Quasiparticle Wave Packets}
\label{sec:quasiparticle}

Wave packets with spatially concentrated energy behave analogously to classical particles and may be characterized by their central wavenumber,
\begin{equation}
\label{eq:center_wavenumber}
\bm{K} = \int \bm{k} |f(\bm{k})|^2 d^3k,
\end{equation}
and their central frequency,
\begin{equation}
\label{eq:center_frequency}
\Omega = \int \omega(\bm{k}) |f(\bm{k})|^2 d^3k.
\end{equation}

Setting $\bm{q} = \bm{k} - \bm{K}$, we observe that $\displaystyle\int \bm{q} |f|^2 d^3q = 0$ follows from the definition of the central wavenumber. Taylor-expanding the dispersion relation $\omega(\bm{k}) = \sqrt{c^2 k^2 + \omega_0^2}$ about $\bm{K}$ and retaining only the first-order term, we obtain $\omega(\bm{K}+\bm{q}) = \omega(\bm{K}) + \bm{v}_g(\bm{K})\cdot\bm{q}$, where $\bm{v}_g = \partial\omega/\partial\bm{k}$ denotes the group velocity. Substituting this expansion into the expression for the central frequency, we find that the integral of the linear term vanishes. Neglecting higher-order terms, we obtain $\Omega \approx \omega(\bm{K}) = \sqrt{c^2 K^2 + \omega_0^2}$. Squaring both sides of this relation yields the dispersion relation satisfied by the wave-packet center,
\begin{equation}
\label{eq:center_dispersion}
\Omega^2 \approx c^2 K^2 + \omega_0^2.
\end{equation}
The group velocity of the wave packet reads $\bm{v}_g = \partial \omega/ \partial \bm{k} \big|_{\bm{k}=\bm{K}} = c^2 \bm{K}/\Omega \leq c$.

It follows that the energy carried by the wave packet propagates at the group velocity, whose magnitude cannot exceed the transverse wave speed $c$ of the SM. This implies that the transverse wave speed $c$ serves not only as the propagation speed of wave phases but also as the limiting speed for energy and information transfer.

We introduce a constant $\hbar$ with the dimensions of action to convert frequency and wavenumber into energy and momentum according to $E = \hbar\Omega$ and $\bm{p} = \hbar\bm{K}$, where $m = \hbar\omega_0/c^2$ denotes the effective mass. This mapping establishes a correspondence at the dimensional level, rendering the dispersion relation of wave packets mathematically isomorphic to the energy-momentum relation of a standard relativistic free particle.

\subsection{Wave Packets with Intrinsic Degrees of Freedom}
\label{sec:multicomponent}

Consider a spatially localized wave-packet system \(S\). Suppose there exist \(N\) normal modes \(\{\psi_a(\bm{x},t), a=1,2,\dots,N\}\) forming a complete basis for its state space, and every \(\psi_a\) satisfies the wave equation \eqref{eq:klein_gordon_scalar}. Assemble these \(N\) modes into an \(N\)-dimensional complex vector
\begin{equation}
\label{eq:mode_vector}
\Psi(\bm{x},t) = \big( \psi_1, \psi_2, \dots, \psi_N \big)^T \in \mathbb{C}^N,
\end{equation}
which is defined as the multicomponent wavefunction of system \(S\).

It should be emphasized that these \(N\) modes generally span an abstract \(N\)-dimensional intrinsic state space, rather than \(N\) polarization directions confined within 3-dimensional physical space. Each component \(\psi_a(\bm{x},t)\) remains a scalar field in physical space (obeying the scalar wave equation), while the index \(a\) labels distinct basis vectors of intrinsic degrees of freedom. Its physical origin may consist of \(N\) independent normal modes within a single spatial polarization direction, or other forms of intrinsic excitations. For the purpose of this paper, the concrete physical realization is irrelevant; what matters is that these \(N\) modes mathematically constitute a linearly independent state space. In a dissipation-free elastic SM, energy conservation ensures that localized excitations do not decay via dissipation, and modes with different wavenumbers share an identical dispersion relation within a homogeneous and isotropic SM.

Since each mode obeys the identical wave equation \eqref{eq:klein_gordon_scalar}, the dynamical equation for system \(S\) can be written as
\begin{equation}
\label{eq:multicomponent_dynamics}
L \Psi \equiv L_0 I_{N \times N} \Psi = 0,
\end{equation}
where \(L_0\) stands for the scalar wave differential operator. As \(L = L_0 I_{N \times N}\) commutes with any constant matrix \(U\), for any constant unitary matrix \(U \in U(N)\) we have \(U L = L U\). Therefore, if \(\Psi\) is a solution, \(U\Psi\) is also a solution, indicating that the solution space of the system carries a representation of the \(U(N)\) group.

Analogous to the fact that global phase transformations of scalar wavefunctions correspond to the choice of reference orientation on the complex plane, distinct group elements \(U \in U(N)\) correspond to alternative choices of reference basis in the \(N\)-dimensional intrinsic state space. Different reference bases yield different component expressions yet describe the identical physical state, which constitutes the physical foundation for the system to possess intrinsic \(U(N)\) symmetry.

Global phase transformations belong to the Abelian subgroup \(U(1)\) of \(U(N)\); such transformations leave the relative amplitudes and relative phases between all components unchanged and carry no physical information regarding intrinsic states. After removing this redundant symmetry, the effective symmetry group reduces to \(SU(N)\). The \(SU(N)\) group governs the symmetries characterizing the relations among all oscillatory modes composing the wave packet, and we refer to \(SU(N)\) as the \textit{intrinsic symmetry} of the wave packet.

\subsection{Identity Invariance of Wave Packets}
\label{sec:identity}

For wave packets endowed with $SU(N)$ intrinsic symmetry, an arbitrary group transformation $U \in SU(N)$ acting as $\Psi \to U\Psi$ corresponds to a reselection of the reference basis in the intrinsic state space. Such rotations of intrinsic coordinate frames leave the physical state of the wave packet unaltered. All wavefunctions related by $SU(N)$ transformations are therefore physically equivalent, which motivates the introduction of equivalence classes defined as
\begin{equation}
\label{eq:identity_class}
[\Psi] = \{ U\Psi \mid U \in SU(N) \},
\end{equation}
where wavefunctions belonging to the set are regarded as physically identical. The  equivalence class $[\Psi]$ thereby labels the identity of the wave packet.

Two crucial corollaries follow directly from the notion of  equivalence class. First, all physical observables are functions on equivalence class and hence invariant under $SU(N)$ transformations. Since every wavefunction belonging to $[\Psi]$ describes a single unique physical state, any physical observable must assume the same numerical value for all elements of the same equivalence class. The energy density $\Psi^\dagger \Psi$ of the wave packet constitutes a fundamental invariant; more generally, any $SU(N)$-invariant functional constructed from $\Psi$ and its derivatives qualifies as a physical observable. Second, the  equivalence class $[\Psi]$ of intrinsic states remains unchanged throughout the dynamical evolution of the wave packet. As a wave packet propagates through spacetime, its physical identity must persist intact; otherwise the wave packet would undergo decomposition or loss of identity. We term this core constraint the requirement of identity invariance for wave packets.

The requirement of identity invariance is a dynamical constraint rather than a mere definitional convention, imposing a stringent restriction on the evolution of wave packets. A shift in the  equivalence class of intrinsic states would signify a modification of the intrinsic symmetric structure of the wave packet itself, amounting to a change in its fundamental identity. This constraint of identity invariance must hold unconditionally for stable wave packets interpreted as elementary particles.

When a wave packet propagates through an inhomogeneous background medium, the reference basis for intrinsic states can be chosen independently at distinct spacetime points. Interactions between the background field and the wave packet manifest themselves as rearrangements of intrinsic states confined strictly within the  equivalence class, without altering the  equivalence class itself. If the  equivalence class must be preserved globally across all spacetime locations, the background medium must furnish a mechanism to reconcile discrepancies between locally chosen reference bases at separate points. This mechanism will be elaborated in Sec.\ref{sec:gauge}: the SM spontaneously generates gauge fields in response to this global consistency demand, thereby ensuring that the identity of the wave packet is sustained over the entire spacetime domain. From this perspective, gauge fields are not artificially devised mathematical constructs, but rather an unavoidable response of the medium arising directly from the dynamical postulate of identity invariance.

\section{Gauge Structures in Generalized Continuum Mechanics}
\label{sec:continuum}

As outlined in Sec.\ref{sec:introduction}, the mathematical apparatus of gauge theory is deeply embedded within classical GCM. The $SO(3)$ frame field together with its associated connection and curvature constitutes one special yet vital instance. This structure furnishes an isomorphic foundation for the subsequent discussion of $SU(N)$ gauge fields and $SO(1,3)$ gravitational frame fields. Geometric quantities defined in elastic media---torsion rate and curvature tensor---share identical algebraic structures with their counterparts in gauge field theory, namely the connection and the field strength tensor.

Generalized continua endowed with microrotational degrees of freedom, such as Cosserat media~\cite{Cosserat1909} and micropolar elastic solids~\cite{Eringen1968,Eringen1999}, are distinguished by a core feature: each material point carries not only translational degrees of freedom but also independent rotational degrees of freedom. A local frame field $R(x) \in SO(3)$ is attached to every material point to characterize the local orientation of the medium in three-dimensional space. The frame field $R(x)$ arises from the geometric description of material microstructures, rather than from \textit{a priori} symmetry postulates---a critical distinction from the introduction of gauge fields in high-energy physics.

Spatial variations of the frame are encoded in the gradient $\partial_i R$; however, $R(x)$ itself depends on the choice of global coordinate system and hence does not qualify as a geometric invariant. To resolve this, we define the relative torsion rate
\begin{equation}
\label{eq:classical_torsion}
\Gamma_i = R^T \partial_i R,
\end{equation}
an antisymmetric matrix quantifying the relative rotation rate of frames along the $i$-th spatial direction, that is, the net rotation of the local frame when moving from one material point to an adjacent point. The relative torsion rate $\Gamma_i$ constitutes a physical quantity independent of global rotations and takes values in the Lie algebra $so(3)$, expandable in terms of the generating basis elements of $so(3)$. Furthermore, the net rotation accumulated by transporting the frame field around an infinitesimal closed loop is measured by the curvature tensor
\begin{equation}
\label{eq:classical_curvature}
\Omega_{ij} = \partial_i \Gamma_j - \partial_j \Gamma_i + [\Gamma_i, \Gamma_j].
\end{equation}
Within geometric defect theory, nonvanishing $\Omega_{ij}$ is interpreted as a measure of disclination density, and the curvature tensor corresponds to the areal density of Frank vectors~\cite{Katanaev2005}.

Defects induced by intrinsic curvature are generically accompanied by elastic stresses, imposing an energetic cost on the system. Under the small-deformation approximation, the energy density is typically constructed as a quadratic form in the curvature tensor~\cite{Eringen1968,Nowacki1986,Eringen1999,Katanaev2005}, corresponding to the lowest-order nonvanishing contribution. For homogeneous isotropic elastic media, this energy density reads
\begin{equation}
\label{eq:classical_energy_density}
\mathcal{L}_{\text{int}} = \frac{K}{2} \operatorname{Tr}(\Omega_{ij} \Omega^{ij}),
\end{equation}
where $K$ denotes the generalized elastic modulus whose numerical value is determined by intrinsic medium properties, and $\Omega^{ij}$ is obtained by raising indices via the Euclidean metric $\delta^{ij}$. This expression represents the minimal admissible energy functional for isotropic media. In gauge-theoretic formulations of defect mechanics, this curvature-based energy density mirrors the form of the Yang-Mills action, and the equilibrium equations derived via variational calculus are mathematically isomorphic to Yang-Mills-type field equations~\cite{Kadic1983,Lazar2000,Peshkov2019}. The constructive procedure for building a quadratic energy density from the curvature tensor is formally identical to the construction of the Yang-Mills action. Physically, however, $K$ here acts as an elastic modulus governed by medium characteristics, whereas the coupling constant $g$ in Yang-Mills theory originates from an entirely distinct physical mechanism.

The frame field $R(x)$, the associated torsion rate $\Gamma_i$, and the curvature $\Omega_{ij}$ introduced above constitute the fundamental building blocks of the $SO(3)$ gauge structure within classical GCM, and realize an $SO(3)$ frame bundle over the spatial manifold. In subsequent sections, we establish two corresponding mappings based upon this formalism: first, extending spatial indices from three-dimensional Euclidean space to four-dimensional Minkowski-type spacetime; second, mapping the real-valued $SO(3)$ intrinsic frame fields of micropolar media onto the complex $SU(N)$ intrinsic frame fields of wave packets. Across both correspondences, the mathematical structures---frame fields, connections, and curvature---remain identical, yet the physical carriers differ at each hierarchical level: three-dimensional spatial frames are superseded by four-dimensional spacetime frames, and rotational degrees of freedom in elastic media are replaced by the intrinsic symmetries of wave packets.

\section{Emergence of Spacetime Structure}
\label{sec:spacetime}

Under the conventionalist time-synchronization scheme, this section demonstrates how dynamical relations among wave packets within a homogeneous SM map isomorphically onto the geometric structure of Minkowski-type spacetime. This mapping constitutes a kinematic isomorphism. Our arguments illustrate that the mathematical formalism of SR can be reconstructed isomorphically upon the SM, offering a viable interpretive route wherein spacetime geometry arises jointly from wave-packet dynamics and observational conventions. The spacetime we perceive is not a geometric arena prescribed \textit{a priori}, but an effective structure constructed from the transmission rules of physical signals. 

\subsection{The Unity of Spacetime}
\label{sec:spacetime_unity}

Velocity is defined as the ratio of spatial displacement to time interval; therefore, referring to velocity prior to independently defining space and time appears logically inconsistent. Nevertheless, separate \textit{a priori} definitions of space and time entrap reasoning within the philosophical dualism of decoupled space and time---a tension epitomized by Zeno's paradoxes, including the Achilles--Tortoise and Arrow paradoxes. Spacetime must therefore be conceptualized and described as an inseparable unity, which mathematically demands a fundamental physical quantity linking spatial and temporal measures. Velocity fulfills this epistemologically foundational role and must be presupposed in any operational construction of spacetime concepts.

SR adopts the speed of light in vacuum---the propagation velocity of electromagnetic waves---as the physical signal for perceiving spacetime, postulates this speed as a universal constant, and employs it to operationally define time and subsequently space, thereby unifying spacetime at the operational level. In this sense, the speed of light is not a derived consequence of more primitive axioms, but rather a conventional operational benchmark for calibrating spacetime measurements. Fixing the speed of light as a universal constant to define space and time ensures that spacetime transforms as an integrated whole, rather than permitting space and time to be treated as independent entities under Galilean transformations; this constitutes the physical foundation of Lorentz transformations. Consequently, spatial and temporal quantities can be defined only locally, rather than globally.

The SM might appear to constitute an absolute rest frame, yet this is not the case. As shown below, adopting the convention that the transverse wave speed of the SM is isotropic renders this frame no more privileged than any other inertial frame. We define \textit{inertial frames} as reference systems at rest or in uniform motion relative to the SM. The SM is postulated as a homogeneous, isotropic continuous substance, a supposition formulated in the inertial frame at rest relative to the SM.

This setup raises an operational question: what physical signal velocity should be adopted as the presupposed benchmark for the joint definition of space and time?

\subsection{Emergence of the Spacetime Metric}
\label{sec:metric_emergence}

For observers at rest relative to the SM, space is homogeneous and isotropic, and transverse wave excitations propagate in all directions at the identical speed $c = \sqrt{\mu/\rho}$, which also sets the upper bound for energy and information transfer. For observers in uniform motion relative to the SM, it is operationally meaningless to inquire whether transverse waves propagate at equal speeds along and against the direction of motion prior to the establishment of a consistent time standard. Velocity measurement inherently relies on well-defined time intervals, the definition of which in turn presupposes a synchronization convention~\cite{Reichenbach1958}.

We adopt an operational scheme isomorphic to that of SR~\cite{Einstein1905b}, conventionally postulating the transverse wave speed $c$ of the SM as a universal constant. Explicitly, two-way synchronization via transverse wave signals defines simultaneity within any inertial frame. Let a signal be emitted from point $A$ at instant $t_A$, reflected at point $B$, and returned to $A$ at instant $t_A'$. The synchronization condition for the clocks at $B$ and $A$ reads
\begin{equation}
t_B = t_A +(t_A' - t_A)/2.
\end{equation}

Spatial distance is defined as the product of signal speed and time interval---an operational principle identical to the formal definition of the metre in the modern International System of Units~\cite{BIPM2019}. The transverse wave speed $c$ is an intrinsic material parameter of the SM and serves as the upper limit of group velocities for quasiparticle wave packets; within our framework, it assumes precisely the functional role occupied by the speed of light in vacuum in SR. This conventional choice nECsarily yields Lorentz transformations and an emergent Minkowski-type spacetime geometry, with the set of all Lorentz transformations constituting the Lorentz group $SO(1,3)$. The spacetime interval
\begin{equation}
ds^2 = -c^2 dt^2 + d\bm{x}^2
\end{equation}
remains invariant under arbitrary coordinate transformations, and the Minkowski metric $\eta_{\mu\nu} = \mathrm{diag}(-1,1,1,1)$ emerges naturally. In subsequent sections, we employ this emergent spacetime structure and introduce four-dimensional coordinates $x^{\mu}$ ($\mu = 0,1,2,3$; $x^0 = c t$).

By conventionally stipulating the one-way transverse signal speed to be the isotropic constant $c$ in all inertial frames, and employing this constant to define time, all kinematic effects arising from motion relative to the SM are fully absorbed into the mathematical structure of spacetime definitions at the operational level. The transverse wave speed---an intrinsic property of the SM---is elevated to the operational benchmark for calibrating spacetime measurements. It follows that emergent Minkowski-type spacetime is not a physically derived substance, but rather an effective geometric structure isomorphically encoded through conventionalist measurement protocols. The SM recedes into the background as the fundamental substrate supporting wave-packet excitations, while its intrinsic properties---homogeneity, isotropy, and transverse wave speed---are fully subsumed within the spacetime metric and cannot be directly detected through measurements employing transverse wave signals.

This furnishes an alternative interpretive lens for the null result of the Michelson-Morley experiment: the temporal and spatial standards employed in the experiment are themselves defined by light signals, and this operational definition fully absorbs any hypothetical medium effects, rendering them undetectable through optical measurements. Observers themselves are composite wave-packet excitations, and their entire measurement framework internalizes medium-related corrections. The SM therefore does not manifest as an observable background arena; it exists as a necessary precondition enabling geometric descriptions of wave-packet dynamics.

Inertial frames are identified as physical reference systems anchored to material bodies, rather than as \textit{a priori} absolute space. The convention that the transverse wave speed remains invariant in all inertial frames is consistent precisely because all wave packets correspond to transverse excitations of the SM, with $c$ constituting an intrinsic medium property. Spacetime metrics are well-defined only within local reference frames, and the metrics of distinct inertial frames are interrelated by Lorentz transformations---an unavoidable consequence of the conventionalist operational strategy.

Notably, Einstein~\cite{Einstein1920} argued that a physically meaningful aether cannot possess traceable degrees of freedom characteristic of ponderable matter, nor can the concept of motion be attributed to it. The homogeneous SM introduced herein exists solely as a supporting substrate for wave-packet excitations and cannot be sensed by transverse waves or their associated wave packets, thereby precluding any consistent assignment of motion to the medium itself. Its intrinsic characteristics are encoded within the spacetime metric, and it does not constitute a locally labelable absolute reference frame.


\section{Emergence of Gauge Fields}
\label{sec:gauge}

This section applies the fiber-bundle geometric structure introduced in the context of Newtonian absolute spacetime in Sec.\ref{sec:continuum} to the localization of intrinsic symmetries carried by wave packets on Minkowski-type spacetime, establishing a mathematical isomorphism. Specifically, we replace the frame field $R(x) \in SO(3)$ by local transformations $U(x) \in SU(N)$ of the intrinsic symmetry group, and substitute the spatial derivative $\partial_i$ with the four-dimensional spacetime derivative $\partial_\mu$. The connection defined on the spatial-rotation fiber bundle thereby corresponds to the gauge field $A_\mu$, while the associated curvature maps onto the gauge field strength $F_{\mu\nu}$.

\subsection{Introduction of Local Symmetry}
\label{sec:local_symmetry}

On Minkowski-type spacetime, the identity of a wave packet $\Psi(x)$ is characterized by the  equivalence class of intrinsic states $[\Psi]$ defined in Eq.~\eqref{eq:identity_class}, with $x$ denoting the four-dimensional spacetime coordinates $x^{\mu}$. For a homogeneous and static background of the SM, global transformations $\Psi \to U\Psi$ with constant $U \in SU(N)$ leave the  equivalence class invariant; such a constant $U$ corresponds to a global symmetry.

When the reference basis for intrinsic states may be chosen independently at distinct spacetime points, global transformations must be promoted to spacetime-dependent local transformations
\begin{equation}
\label{eq:local_transform}
\Psi(x) \to \tilde{\Psi}(x) = U(x) \Psi(x), \quad U(x) \in SU(N).
\end{equation}
The physical foundation of this generalization lies in the natural extension of the requirement of identity invariance (identity invariance) to the local level. The global identity of a wave packet is labeled by the  equivalence class $[\Psi] = \{ \tilde{\Psi}(x) \}$, which remains unaffected by local choices of reference basis. If the requirement of invariant  equivalence class must hold globally, a consistent connection rule is required to link wave-packet states across different spacetime points; this furnishes the physical motivation for introducing gauge fields as connections. We analyze this issue starting from the wave-packet dynamical equation
\begin{equation}
\label{eq:free_dynamics}
H(\partial_\mu) \Psi = 0.
\end{equation}

Directly substituting the  equivalence class wavefunction $\tilde{\Psi}(x)$ from \eqref{eq:local_transform} for $\Psi(x)$ in the dynamical equation \eqref{eq:free_dynamics} destroys the form invariance of wave-packet dynamics, since derivative terms $\partial_\mu \Psi$ yield additional terms under local transformations. We adopt an alternative approach: instead of locally transforming the wavefunction $\Psi(x)$ directly, we apply a local unitary transformation $U(x) \in SU(N)$ to the entire dynamical equation \eqref{eq:free_dynamics}, while inserting the resolution of the identity $I = U^\dagger \cdot U$. Rearranging Eq.~\eqref{eq:free_dynamics} yields
\begin{equation}
\label{eq:u_transform_dynamics}
U H(\partial_\mu) U^\dagger \cdot U \Psi = 0.
\end{equation}
This relation embodies the dual effect of background variations on both physical states and dynamical operators, implying that dynamical operators must be adjusted accordingly whenever the intrinsic state of a wave packet is modified.

The transformation rule for partial differential operators reads
\begin{equation}
\label{eq:derivative_transform}
U \partial_\mu U^\dagger = \partial_\mu + U(\partial_\mu U^\dagger) .
\end{equation}
We introduce a coupling constant $g$ and define the gauge field $A_\mu$ via the identity
\begin{equation}
\label{eq:gauge_field_definition}
i g A_\mu =  U(\partial_\mu U^\dagger) ,
\end{equation}
which recasts the transformed derivative as
\begin{equation}
\label{eq:derivative_with_gauge}
U \partial_\mu U^\dagger = \partial_\mu + i g A_\mu.
\end{equation}

Since $U(\partial_\mu U^\dagger)$ is an anti-Hermitian matrix, the gauge field $A_\mu$ takes values in the Lie algebra $\mathfrak{su}(N)$ of $SU(N)$, and admits the expansion
\begin{equation}
\label{eq:gauge_field_expansion}
A_\mu = \sum_{a=1}^{N^2-1} A_\mu^a T_a,
\end{equation}
where $T_a$ denote the generators of $su(N)$.

Under the constraint that the wave-packet identity is preserved, the term $i g A_\mu$ quantifies the necessary background response compensating for discrepancies arising from independent local selections of the reference basis for intrinsic states. The explicit form of this response is determined by the intrinsic $SU(N)$ symmetry of the wave packet, and encodes the interaction mechanism between wave packets and the SM background. In other words, the gauge field $A_\mu$ represents the differential response of the SM to distinct intrinsic symmetry configurations. Mathematically, it serves as the connection representing the spacetime gradient of the intrinsic frame $U(x)$ along the $\mu$-direction, and shares the same geometric origin as the relative torsion rate defined in Eq.~\eqref{eq:classical_torsion} of Sec.\ref{sec:continuum}. The two quantities act on distinct fiber bundles: the latter describes geometric responses of material microstructures, while the former corresponds to gauge responses associated with the intrinsic degrees of freedom of wave packets.

Thus, the introduction of gauge fields $A_\mu$ stems from the necessary response of the physical background to locally varying reference bases for intrinsic states. Independent local basis choices at separate spacetime points create discontinuities requiring geometric connections between distinct points, and $A_\mu$ constitutes the mathematical object encoding such connections, thereby enabling the global preservation of wave-packet identity.

An analogous transformation law applies to higher-order differential operators
\begin{equation}
\label{eq:operator_transform}
U H(\partial_\mu) U^\dagger = H(\partial_\mu + i g A_\mu).
\end{equation}

Taking into account the background response of the SM under local unitary transformations of wave packets, the dynamical equation \eqref{eq:u_transform_dynamics} becomes
\begin{equation}
\label{eq:transformed_dynamics}
H(\partial_\mu + i g A_\mu) (U \Psi) = 0.
\end{equation}

We define the covariant derivative
\begin{equation}
\label{eq:covariant_derivative}
D_\mu = \partial_\mu + i g A_\mu,
\end{equation}
such that the dynamical equation simplifies to
\begin{equation}
\label{eq:covariant_dynamics}
H(D_\mu) \tilde{\Psi} = 0,
\end{equation}
which has the same mathematical form as Eq.~\eqref{eq:free_dynamics}, with ordinary partial derivatives replaced by covariant derivatives.

Within the dynamical equation incorporating the background response of the SM, the physical identity of wave packets is intrinsically labeled by  equivalence class; that is, $\tilde{\Psi}$ stands for an arbitrary element of the class $[\Psi]$. The only required modification to Eq.~\eqref{eq:free_dynamics} is the substitution of the ordinary partial derivative $\partial_\mu$ with the covariant derivative $D_\mu$. This substitution compensates for the medium response term $U(\partial_\mu U^\dagger)$ arising from local discrepancies in intrinsic reference bases, and thereby ensures form invariance of the dynamical description over the entire  equivalence class. This formulation is mathematically isomorphic to standard quantum field theory, where the gauge redundancy structure of quantum states maps one-to-one onto the equivalence-class structure developed here.

From the definitions of gauge fields and covariant derivatives, their transformation laws under local gauge transformations follow directly,
\begin{align}
\label{eq:gauge_transform_A}
A_\mu &\to U A_\mu U^\dagger + \frac{1}{i g} U(\partial_\mu U^\dagger),\\
\label{eq:covariant_derivative_transform}
D_\mu &\to U D_\mu U^\dagger.
\end{align}
This transformation behavior further implies that the covariant derivative $D_\mu \Psi$ transforms in the same manner as the wavefunction $\Psi$ itself.


\subsection{Geometric Structure of Gauge Fields}
\label{sec:gauge_geometry}

Discrepancies arising from independent local choices of reference bases for wave-packet intrinsic states are equivalent to inducing corresponding deformations in the medium background. The introduction of gauge fields $A_\mu$ establishes connections between intrinsic states defined at distinct spacetime points. The global consistency of such connections, however, hinges on whether nontrivial modifications of intrinsic states emerge after parallel transport along closed loops---a property quantified by the curvature of the gauge field, namely the field strength tensor. Within the framework of this paper, the field strength tensor is not merely a geometric quantity; it simultaneously quantifies the magnitude of deformations sustained by the SM in response to local rearrangements of wave-packet intrinsic states. More pronounced deformations yield larger field strength values and impose a greater energetic cost on the medium.

Generalizing the definition of the curvature tensor $\Omega_{ij}$ in Eq.~\eqref{eq:classical_curvature} to Minkowski tpye spacetime, we define the field strength tensor associated with the gauge field introduced in Eq.~\eqref{eq:gauge_field_definition} as
\begin{equation}
\label{eq:field_strength_def}
F_{\mu\nu} = \partial_\mu A_\nu - \partial_\nu A_\mu + i g [A_\mu, A_\nu],
\end{equation}
which measures the nonintegrability of the twist in the intrinsic frame field $U(x) \in SU(N)$. Under gauge transformations, the field strength tensor transforms as
\begin{equation}
\label{eq:gauge_transform_F}
F_{\mu\nu} \to U F_{\mu\nu} U^\dagger,
\end{equation}
demonstrating that the geometric interpretation of $F_{\mu\nu}$ is independent of the choice of reference basis.

The introduction of covariant derivatives guarantees the form invariance of the dynamical equation~\eqref{eq:covariant_dynamics}; yet this formalism does not address the energetic cost incurred by the SM in sustaining gauge fields. The fundamental guiding principle here is the minimization of the total energetic expenditure. This condition essentially demands that the connection field $A_\mu$ arise from the optimal compensatory response of the SM to counteract deformation effects, thereby preserving the invariant identity of wave packets at all spatial locations they traverse.

The energetic cost of this medium response is quantified by a quadratic form in the field strength tensor $F_{\mu\nu}$. At the lowest-order approximation, this represents the unique energy functional simultaneously satisfying gauge invariance, isotropy, and linearity in the weak-field limit. Higher-order terms such as $\operatorname{Tr}(F_{\mu\nu}F^{\mu\nu})^2$ or $\operatorname{Tr}(F_{\mu\nu}F^{\nu\rho}F_{\rho\mu})$ are formally compatible with gauge invariance, yet are suppressed in the weak-field regime. Accordingly, the Yang-Mills action
\begin{equation}
\label{eq:yang_mills_action}
S_{\text{YM}} = -\frac{1}{2} \int d^4x \, \operatorname{Tr}(F_{\mu\nu} F^{\mu\nu}),
\end{equation}
constitutes the minimal admissible functional subject to the two core constraints of wave-packet identity invariance and minimal energetic cost. Variation of this action yields the Yang-Mills type field equations
\begin{equation}
\label{eq:yang_mills_equation}
D_\mu F^{\mu\nu} = 0.
\end{equation}

This framework is mathematically isomorphic to conventional Yang-Mills theory: the gauge field $A_\mu$ acts as a connection on a principal fiber bundle, while its associated field strength $F_{\mu\nu}$ corresponds to the bundle curvature. The critical distinction lies in the physical interpretation. In standard Yang-Mills theory, gauge fields are postulated as fundamental dynamical degrees of freedom. By contrast, within the present formalism, $A_\mu$ emerges as an unavoidable background response enforced by the requirement of identity invariance, encoding the mutual feedback between local deformations of the SM and locally chosen reference bases for intrinsic states. Different values of $N$ correspond to distinct intrinsic symmetries of wave packets. As is evident from Eq.~\eqref{eq:gauge_field_expansion}, variations in intrinsic symmetry dictate distinct response modes of the SM to wave-packet motion---that is, the specific interaction forms encoded by the components $A_\mu^a$.

\section{Curved Spacetime as the Geometric Encoding of an Inhomogeneous Substrate Medium}
\label{sec:gravity_emergence}

Section~\ref{sec:spacetime} addressed the case of a homogeneous SM. Via conventionalist measurement protocols, the homogeneous medium gives rise to Minkowski-type spacetime, with coordinate transformations linked by global Lorentz mappings. wave-packet excitations introduce local deformations into the SM through their energy--momentum tensor $T_{\mu\nu}$. More severe deformations correspond to greater inhomogeneity of the medium distribution, manifested as spatial variations in the wave speed $c(\bm{x})$. Under such circumstances, the conventionalist scheme combining two-way signal synchronization with the convention of an isotropic one-way signal velocity is valid only locally, yielding a geometric structure with independent local Lorentz frames at each spacetime point. This mirrors the local freedom of reference-basis selection encountered in the preceding gauge-field discussion and inevitably leads to a description in terms of curved spacetime.

\subsection{Fiber-Bundle Formalism}

The $SO(3)$ frame field of GCM in Sec.\ref{sec:continuum} constitutes an explicit realization of a frame bundle, whereas the $SU(N)$ principal-bundle structure for wave-packet intrinsic symmetries in Sec.\ref{sec:gauge} follows the same mathematical pattern. Since inhomogeneous SM distributions necessarily require curved spacetime geometry, the same formal framework remains applicable: a local frame $e^a_\mu(x)$ is attached to each spacetime point $x$. These local frames possess the structure group $SO(1,3)$, corresponding to the group of local Lorentz transformations. The metric $g_{\mu\nu} = \eta_{ab}e^a_\mu e^b_\nu$ is a derived quantity constructed from frame fields. Correlations between distinct local frames are governed by the spin connection $\omega^{ab}_\mu$, and the nonintegrability of frames transported around closed loops is quantified by the curvature tensor $R^{ab}_{\mu\nu}$. The full physics of an inhomogeneous medium can be encoded in the language of Riemannian geometry---a fundamental isomorphic mapping.

\subsection{Elastic Response of the Substrate Medium}

Parallel transport defined via fiber bundles furnishes a constructive procedure for wave-packet dynamics in an inhomogeneous SM: we generalize the variational principle for the three-dimensional spatial frame field $R(x) \in SO(3)$ from Sec.\ref{sec:continuum} to four-dimensional spacetime frame fields $e_\mu^a(x)$.

Local deformations of the SM at each spacetime point $x$ are parameterized by the frame field $e_\mu^a(x)$. Under conventionalist operational rules, these deformations emerge as the curved spacetime metric $g_{\mu\nu}(x) = \eta_{ab} e_\mu^a(x) e_\nu^b(x)$. The elastic response of the medium to deformation is measured by the first-order gradient of frame fields, $\partial_\mu e_\nu^a$. For elastic media, equilibrium configurations correspond to variational extrema of the free-energy functional. We extend this variational principle to spacetime frame fields: the dynamical response of the SM to inhomogeneous deformations coincides with configurations extremizing the total action.

A physically plausible ansatz is that, within the low-order long-wavelength approximation, the coordinate-invariant self-interaction free-energy functional assumes a curvature-scalar form
\begin{equation}
\label{eq:gravity_action}
S_{\text{grav}}[e] = \frac{1}{16\pi G} \int d^4x \, \det(e) \, R(e, \partial e) + S_{\text{matter}}(e, \Psi),
\end{equation}
where $\det(e) = \det(e_\mu^a)$, $R = e^\mu_a e^\nu_b R^{ab}_{\mu\nu}$ denotes the curvature scalar constructed from frame fields and their first derivatives, and $G$ is an effective coupling constant associated with the elastic modulus of the SM. Within this framework, $G$ represents the collective stiffness coefficient governing the resistance of the SM to deformation, with its magnitude fixed by experiment. The term $S_{\text{matter}}$ denotes the matter-field action for wave-packet excitations introduced in Sec.\ref{sec:wavepacket}.

This setup differs crucially from the quadratic curvature energy density $\operatorname{Tr}(\Omega_{ij}\Omega^{ij})$ of Sec.\ref{sec:continuum}, whose variation yields fourth-order field equations describing static disclination defects in three-dimensional elastic solids. By contrast, the curvature scalar $R$ built from four-dimensional spacetime frame fields already contains second derivatives of the frame variables; retaining only the linear curvature term naturally produces second-order field equations, consistent with the principle of the lowest-order nontrivial contribution in four dimensions. Functionals proportional to $R^2$ would generate fourth-order field equations associated with nonlocal elastic responses or higher-gradient theories~\cite{Katanaev2005}, and are discarded in the long-wavelength limit.

Taking the variational derivative of the gravitational action with respect to the frame fields and imposing $\delta S_{\text{grav}} / \delta e_\mu^a = 0$, in the torsion-free approximation, frame-field variation is equivalent to metric variation; yet the frame-based formulation more transparently encodes the microstructural description of the SM. The variational calculation follows a standard procedure mathematically identical to the derivation of Einstein's field equations from the Einstein--Hilbert action in GR, with the sole difference that the fundamental dynamical field is the frame $e_\mu^a$ rather than the metric $g_{\mu\nu}$. The resulting variational identity reads
\begin{equation}
\label{eq:einstein_derivation}
G_{\mu\nu} \equiv R_{\mu\nu} - \frac{1}{2} g_{\mu\nu} R = 8\pi G \, T_{\mu\nu},
\end{equation}
where $R_{\mu\nu} = R^{\lambda}_{\mu\lambda\nu}$ is the Ricci tensor and $T_{\mu\nu} = \frac{2}{\sqrt{-g}} \frac{\delta S_{\text{matter}}}{\delta g^{\mu\nu}}$ denotes the energy--momentum tensor of matter fields (wave-packet excitations).

Two key consequences arise from the spatially varying wave speed $c(\bm{x})$ induced by inhomogeneous SM distributions and medium deformations. At the kinematic level, such variations are directly encoded as the curved metric $g_{\mu\nu}(x)$ via conventionalist measurement rules. At the dynamical level, medium deformations obey the variational field equation~\eqref{eq:einstein_derivation}. Medium inhomogeneity and curved geometry enter into a mutually constructive relation: medium deformations shape the spatial distribution of matter fields, and those same deformations receive geometric encoding as spacetime metrics through conventionalist operational procedures. From this isomorphic viewpoint, Einstein-type field equations do not require postulation as fundamental axioms; they emerge naturally as the variational condition for the elastic-response free energy of the SM recast in geometric language.

This variational derivation is structurally analogous to the geometric theory of defects~\cite{Katanaev2005}. In defect mechanics, curvature tensors are interpreted as disclination densities, and their variation generates equilibrium equations for continuous defective media. Within the present framework, nontrivial frame-field configurations originate from spatial variations of the SM wave speed $c(\bm{x})$---that is, perturbations exerted by the matter energy--momentum tensor $T_{\mu\nu}$ on the SM---which become geometrically encoded under conventionalist measurement rules; their direct variation yields Einstein-type gravitational field equations.

\subsection{Parallel Structures}

The framework developed in this paper is mathematically isomorphic to GR, yet it rests on a distinct ontological foundation. GR interprets $g_{\mu\nu}$ as an intrinsic property of spacetime itself, with its field equations treated as geometric axioms. By contrast, the present framework views $g_{\mu\nu}$ as the geometric encoding of medium distributions---a joint product of the physical properties of the SM, as embodied in spatially varying wave speeds, and conventionalist operational protocols, namely signal-synchronization schemes. The two sides of the Einstein-type field equation~\eqref{eq:einstein_derivation} thereby acquire distinct physical origins: the left-hand side $G_{\mu\nu}$ provides a kinematic geometric description of medium deformations, while the right-hand side $8\pi G\, T_{\mu\nu}$ represents the dynamical variational response of the medium induced by wave-packet excitations. The consistency of these two components is not imposed as a postulate, but emerges as an inherent self-consistency condition linking the elastic response of the SM to geometric encoding through conventionalist measurement rules.

It is possible to describe wave-packet motion fully by merely positing an \textit{a priori} metric $g_{\mu\nu}$ without invoking the SM or its deformations; this is the standard approach of GR. The difference between the present framework and conventional GR lies not in mathematical predictions, but in the interpretive account of the origin of the metric: the metric is the geometric encoding of the physical characteristics of the SM, rather than a primitive geometric entity. This distinction is not distinguishable by internal theoretical observables; however, the two frameworks provide fundamentally different answers to the core question of why the field equations take this specific form---a difference that constitutes the central significance of the constructive interpretation advanced throughout this work.

A striking structural parallel exists among three theoretical formalisms: the $SO(3)$ frame bundle of GCM, the $SU(N)$ principal bundle of gauge field theory, and the $SO(1,3)$ frame bundle describing curved spacetime. All three share the same three-tier differential-geometric fiber-bundle architecture: frame field or group element, connection, and curvature. Table~\ref{tab:parallel_theories} juxtaposes the core concepts of these three theoretical systems.

\begin{table}[htbp]
\centering
\footnotesize
\caption{Parallel Structures across Three Formalisms: GCM, Gauge Field Theory, and Curved Spacetime}
\label{tab:parallel_theories}
\begin{tabular}{c|c|c|c}
\hline\hline
Concept & GCM & Gauge Field Theory & Curved Spacetime \\
\hline
\makecell{Physical\\Substrate}
  & \makecell{Elastic continuum\\(micropolar medium)} 
  & \makecell{Homogeneous SM} 
  & \makecell{Inhomogeneous SM}  \\
\hline
Fiber Bundle 
  & \makecell{$SO(3)$ frame bundle} 
  & \makecell{$SU(N)$ principal bundle} 
  & \makecell{$SO(1,3)$ frame bundle} \\
Frame Field & $R(x) \in SO(3)$ & $U(x) \in SU(N)$ & $e^a_\mu(x)$ (vierbein) \\
Connection & $\Gamma_i = R^T\partial_i R$ & $ig A_\mu = U(\partial_\mu U^\dagger)$ & $\omega^{ab}_\mu$ (spin connection) \\
Curvature & $\Omega_{ij} = \partial_i\Gamma_j - \partial_j\Gamma_i + [\Gamma_i,\Gamma_j]$ & $F_{\mu\nu} = \partial_\mu A_\nu - \partial_\nu A_\mu + [A_\mu,A_\nu]$ & $R^{ab}_{\mu\nu} = \partial_\mu\omega^{ab}_\nu - \partial_\nu\omega^{ab}_\mu + [\omega_\mu,\omega_\nu]^{ab}$ \\
Metric & Induced $g_{ij} = \delta_{ab}R^a_i(x) R^b_j(x)$ & Minkowski $\eta_{\mu\nu}$ & $g_{\mu\nu} = \eta_{ab}e^a_\mu e^b_\nu$ \\
Action Density & $K\operatorname{Tr}(\Omega_{ij}\Omega^{ij})/2$ & $-\operatorname{Tr}(F_{\mu\nu}F^{\mu\nu})/2$ & $\det(e) \, R /16\pi G $ \\
\hline
\hline
\end{tabular}
\end{table}

The parallelism exhibited in Table~\ref{tab:parallel_theories} strongly suggests that curved spacetime gravitational dynamics can be derived from the same variational principle underlying GCM. Each of the three action functionals corresponds to the lowest-order nontrivial term compatible with its respective gauge symmetry and differential-order constraints; their variational extremization yields the elastic equilibrium equations, Yang-Mills-type field equations, and Einstein-type gravitational field equations, respectively.

\section{Discussion and Conclusion}
\label{sec:prospects}

\subsection{Comparisons with Related Frameworks}
\label{sec:comparisons}

We compare several major theoretical paradigms to clarify the similarities and distinctions between the framework proposed in this work and established approaches in the literature.

(i) Emergent gauge fields in Condensed Matter Physics. Chern-Simons descriptions of the quantum Hall effect demonstrate that gauge structures can emerge from underlying microscopic substrates at the quantum scale~\cite{Tsui1982,Laughlin1983}. By contrast, the present framework shows that the full mathematical apparatus of gauge theories can be recovered purely from dynamical constraints on wave packets, even within a classical continuum setting. This indicates that gauge emergence constitutes a far more universal phenomenon, not restricted to quantum many-body effects alone.

(ii) Gauge and gravitational unification in String Theory. String Theory interprets both gauge interactions and gravity as distinct vibrational excitations of one-dimensional strings, thereby furnishing a self-consistent quantum unification scheme~\cite{Polchinski1998}. While the present work shares the objective of String Theory---tracing gauge and gravitational structures back to a more fundamental layer---the two frameworks diverge radically in their foundational premises. String Theory posits a pre-existing higher-dimensional spacetime background and retrieves four-dimensional effective physics via compactification mechanisms. By contrast, the SM is treated herein as the primitive entity; spacetime itself arises as an isomorphic encoded structure, without the need to postulate extra spatial dimensions. String-theoretic unification rests on replacing all elementary particles with distinct string oscillation modes, whereas the unification developed in this paper relies on isomorphic mathematical mappings that trace every geometric and gauge structure to the single physical source of the SM. The present work establishes these correspondences at the classical level, revealing that the shared geometric origin of gauge and gravitational structures can be articulated prior to any quantization procedure. It remains a long-standing unresolved challenge for String Theory to reproduce the complete Standard Model particle spectrum from its fundamental string degrees of freedom~\cite{Marchesano2024}.

(iii) Spin foam models in Loop Quantum Gravity. Loop Quantum Gravity postulates that spacetime is fundamentally composed of discrete quantum-geometric building blocks~\cite{Rovelli2004}. The present framework stands in fundamental tension with this core tenet: we treat spacetime geometry as an effective emergent description of wave-packet dynamics, rather than presupposing discrete quantum spacetime from the outset. If spacetime geometry arises derivatively from the SM, attempting to quantize spacetime geometry in its own right amounts to a category error: the proper candidates for quantization are the microscopic intrinsic degrees of freedom of the underlying medium, not macroscopic geometric fields.

(iv) The emergent gravity paradigm. Sec.~\ref{sec:gravity_emergence} reframes the quantum gravity problem as an inquiry into the collective quantum behavior of the microscopic degrees of freedom of the SM, rather than as the quantization of spacetime geometry. This reorientation aligns with the central insight of emergent gravity research. Jacobson demonstrated from thermodynamic reasoning that Einstein's field equations can be interpreted as an equation of state governing spacetime~\cite{Jacobson1995}. Padmanabhan further argued that gravity emerges as a macroscopic statistical limit of hidden microscopic degrees of freedom, analogous to hydrodynamic or elastic continuum equations~\cite{Padmanabhan2012Ad,Padmanabhan2012Ep,Padmanabhan2014}. Padmanabhan's thermodynamic derivation shares the same conceptual direction as the present framework; the key difference is that our approach identifies the elastic substrate medium as the explicit physical carrier of these microscopic degrees of freedom, whereas Padmanabhan leaves the ontological nature of the underlying entities unspecified. From this vantage point, directly quantizing geometric gravitational fields constitutes a category mistake, just as one would not attempt to quantize pressure fields in fluid dynamics without first quantizing the underlying molecular motion.

(v) Dynamical perspective. Harvey Brown argues that the Lorentz symmetry encoded in Minkowski spacetime is not a primitive geometric fact, but a consequence of the dynamical symmetries inherent in matter field equations such as Maxwell's electrodynamics~\cite{HarveyBrown2005}. On his account, Minkowski geometry does not causally explain relativistic phenomena; instead, it serves merely as a codification of the Lorentz invariance obeyed by fundamental dynamical laws. The present framework shares Brown's core intuition that geometric spacetime structures are encoded quantities, yet the substrates and origins of encoding differ substantially. Whereas Brown traces such encodings to already-established fundamental dynamical laws, the present work extends this paradigm to a deeper foundational level, locating their common origin in the SM and demonstrating how Lorentzian spacetime geometry jointly emerges from SM wave-packet dynamics and conventionalist synchronization protocols. Furthermore, our formalism fully accommodates the emergence of non-Abelian gauge structures, a feature absent from Brown's analysis, which is restricted to relativistic spacetime symmetries alone.

\subsection{Blueprint and Outstanding Technical Problems}
\label{sec:open_issues}

This paper constructs a unified isomorphic derivation of Lorentz symmetry, gauge field structures, and gravitational geometry starting purely from a homogeneous SM at the conceptual level. Nevertheless, several key derivations remain qualitative and demand rigorous technical elaboration in future work.

(i) Fundamental characterization of the SM. Three macroscopic elastic parameters---shear modulus $\mu$, mass density $\rho$, and intrinsic eigenfrequency $\omega_0$---are employed to parameterize the elastic response of the SM throughout this work. These quantities are treated as primitive input assumptions rather than as parameters derivable from an underlying microscopic theory. The continuum approximation adopted for the SM is valid only across the characteristic length scales relevant to wave-packet propagation and dynamics.

(ii) Explicit mapping from normal modes to the Standard Model. The $SU(N)$ intrinsic symmetry arises from dynamical equivalence among $N$ linearly independent normal modes spanning the state space of the wave packet, with no mandatory direct correspondence to a fixed polarization axis in physical space. The present paper establishes only a classical-level isomorphic match to the $SU(3)$ color gauge group of the Standard Model and does not address the quantum-mechanical emergence of particle spectra. Moreover, the framework lacks a natural dynamical selection mechanism that would generate the direct-product gauge structure $SU(3) \times SU(2) \times U(1)$ characteristic of the electroweak and strong interactions.

(iii) Formalism for quantizing medium microscopic degrees of freedom. Sec.\ref{sec:gravity_emergence} redefines quantum gravity as the collective quantum dynamics of the microscopic constituents of the SM rather than as the quantization of spacetime geometry, yet no concrete computational procedure is developed herein to construct a complete quantum theory starting from medium microphysics.

(iv) Potential interpretations of dark matter and dark energy. Intrinsic background energy contained in the SM may generate an effective cosmological constant term in the Einstein-type field equations, offering a novel interpretive route for dark energy and alleviating the severe cosmological constant tension between predictions of quantum field theory vacuum energy and observational bounds. Spatially inhomogeneous self-interaction distributions in the SM could also reproduce the gravitational signatures conventionally attributed to dark matter. Within this physical picture, both dark energy and dark matter correspond to intrinsic bulk properties of the SM itself.

\subsection{Conclusion}
\label{sec:conclusion}

The arguments developed in this paper demonstrate that spacetime geometry, gauge field theories, and gravitational dynamics can receive a unified isomorphic interpretation within a single SM framework.

First, under a conventionalist time-synchronization scheme, the transverse wave speed is stipulated as a universal invariant constant. This operational convention fully absorbs kinematic corrections arising from absolute motion relative to the SM into the definition of spacetime, yielding the isomorphic emergence of Lorentz symmetry and Minkowski-type spacetime from wave-packet dynamics. The null result of the Michelson-Morley experiment receives a novel reinterpretation from this isomorphic perspective: all temporal and spatial measurement standards employed in the experiment are defined via transverse wave signals, and the operational definition inherently encodes all potential medium effects, rendering them undetectable through optical interferometry.

Second, the requirement of identity invariance imposed on the intrinsic degrees of freedom of wave packets drives the emergence of complete gauge structures. The structure group $SU(N)$ governing the internal symmetries of wave packets is determined by the number of linearly independent vibrational normal modes supported by the SM. To preserve invariant  equivalence class of intrinsic states as wave packets propagate through spacetime, the SM spontaneously generates gauge fields as a compensatory response to local basis redefinitions. Mathematically, this response necessitates the introduction of covariant derivatives and the full Yang-Mills action functional.

Third, by generalizing the variational principle of GCM to four-dimensional spacetime frame fields, Einstein-type gravitational field equations are derived from free-energy variations describing the elastic response of the SM to spatially inhomogeneous deformations, within the low-order long-wavelength approximation. The curved spacetime metric $g_{\mu\nu}(x)$ is shown to be the geometric encoding of spatially varying transverse wave speeds $c(\bm{x})$ induced by medium inhomogeneity under conventionalist measurement rules. Curved spacetime geometry is therefore a derived effective description rather than a primitive fundamental postulate of gravitational interactions. This picture simultaneously opens new interpretive pathways for dark energy and dark matter: dark energy may correspond to uniform intrinsic background energy of the SM, while dark matter signatures may originate from geometric distortions induced by spatially uneven self-interactions in the SM.

Fourth, if gravitational spacetime geometry constitutes an emergent effective phenomenon, quantization procedures ought to target the microscopic intrinsic degrees of freedom of the SM rather than macroscopic geometric spacetime fields. This fundamental redefinition reframes the quantum gravity research program as the study of the collective quantum behavior exhibited by the microscopic constituents of the SM.

In summary, the core methodological innovation of this framework lies in rejecting the mathematical formalisms of the Standard Model and GR as primitive axioms, and instead tracing all their structures to isomorphic emergence originating from the single physical substrate medium. Within this unified picture, spacetime geometry, gauge field dynamics, and gravitational field equations are not independent foundational postulates, but tiered effective descriptions arising from a single underlying set of physical rules: the elastic mechanical response of the SM combined with identity-conservation constraints on propagating wave-packet excitations. This constructive reinterpretation reproduces all quantitative predictions of established theories while overhauling their ontological accounts of physical origins. The unification of these three core structures does not rely on group-theoretic amalgamation of separate symmetries, but rather on their shared emergent physical logic rooted in continuum medium dynamics.

Future work will focus on two central open problems: elucidating how the microscopic structure of the SM fixes Standard Model coupling constants and particle mass spectra, and constructing a rigorous quantization pathway from medium microphysics that naturally recovers the well-established effective quantum field theory formalism.

%

%
%
%

\renewcommand\refname{References}
\bibliography{references.bib} 

\clearpage
\phantomsection
\addcontentsline{toc}{section}{Index}
\printindex 
\clearpage

\end{document}